\documentclass[aps,reprint,superscriptaddress,pra,nofootinbib,oneside]{revtex4-1}
\usepackage[utf8]{inputenc}
\usepackage{mathtools}

\usepackage{graphicx}
\usepackage{dcolumn}
\usepackage{bm}
\usepackage{xcolor}
\usepackage{amsmath}
\usepackage{amssymb}
\usepackage{hyperref}

\begin{document}

\title{Fixed-time descriptive statistics underestimate extremes of epidemic curve ensembles}

\date{\today}

\author{Jonas L. Juul}
\email{jlju@dtu.dk}
\affiliation{Department of Applied Mathematics and Computer Science, Technical University of Denmark, Kongens Lyngby, 2800-DK  }

\author{Kaare Gr{\ae}sb{\o}ll}
\affiliation{Department of Applied Mathematics and Computer Science, Technical University of Denmark, Kongens Lyngby, 2800-DK  }

\author{Lasse Engbo Christiansen}
\affiliation{Department of Applied Mathematics and Computer Science, Technical University of Denmark, Kongens Lyngby, 2800-DK  }

\author{Sune Lehmann} 
\email{sljo@dtu.dk}

\affiliation{Department of Applied Mathematics and Computer Science, Technical University of Denmark, Kongens Lyngby, 2800-DK  }
\affiliation{Copenhagen Center for Social Data Science, University of Copenhagen, Copenhagen, 1353-DK}


\newcommand{\ER}{Erd\H{o}s--R\' enyi{ }}


\begin{abstract}
    Across the world, scholars are racing to predict the spread of the novel coronavirus, COVID-19. 
    Such predictions are often pursued by numerically simulating epidemics with a large number of plausible combinations of relevant parameters. 
    It is essential that any forecast of the epidemic trajectory derived from the resulting ensemble of simulated curves is presented with confidence intervals that communicate the uncertainty associated with the forecast.
    Here we argue that the state-of-the-art approach for summarizing ensemble statistics does not capture crucial epidemiological information. 
    In particular, the current approach systematically suppresses information about the projected trajectory peaks. 
    The fundamental problem is that each time step is treated separately in the statistical analysis. 
    We suggest using curve-based descriptive statistics to summarize trajectory ensembles. 
    The results presented allow researchers to report more representative confidence intervals, resulting in more realistic projections of epidemic trajectories and -- in turn -- enable better decision making in the face of the current and future pandemics.
\end{abstract}
\maketitle


Accurately communicating uncertainty is essential when forecasting.
We are currently witnessing an unprecedented effort of scholars across countries and fields, competing to most accurately predict the trajectory of the novel coronavirus, COVID-19, using a plethora of approaches. 
These epidemic forecasts are used by governments aiming to soften the enormous consequences that the pandemic has on economy and health worldwide. 
Particularly crucial to decision makers is information about the overall severity of the epidemic and, in particular, whether local hospitals will be overwhelmed. 
Political decisions to reopen countries and borders are calculated risks.
For that reason, it is of utmost importance that uncertainties and confidence intervals associated with forecasted epidemic trajectories are reliable and well-communicated.

To forecast the trajectory of the novel virus, many researchers simulate the spread of the epidemic using mathematical models. 
Different classes of models are used for this purpose, including deterministic or stochastic compartmental epidemiological models~\cite{holmdahl2020wrong,diekmann2000mathematical} and individual-based models~\cite{ferguson2005strategies}. 
Given initial conditions of the epidemic prevalence in the population and a number of epidemiological and non-epidemiological parameters, these models can simulate hypothetical spreading scenarios. 
In reality, the initial conditions and parameters are not known exactly -- especially not for a new virus like COVID-19. 
For this reason, forecasts are often made by simulating a large number of plausible combinations of these inputs, and then summarizing the resulting \emph{ensemble} of epidemic curves using descriptive statistics. 

Regardless of the kind of model which produced the ensemble of epidemic trajectories, the ensemble is usually summarized using \emph{fixed-time} descriptive statistics, see for example Refs.~\cite{ferguson2005strategies, chinazzi2020effect, yangeight,LosAlamos, JohnHopkins}. 
For each time step, the instantaneous value of curves are ranked from smallest to largest and (possibly weighted) percentiles are computed. 
These percentiles are then used to produce confidence intervals for the forecast on the given time step -- for example, the 50\% least extreme values could be shown by marking the interval making out the 25\textsuperscript{th} to 75\textsuperscript{th} percentiles. 
Below, we show that in the context of forecasting trajectory extremes, however, such fixed-time approaches to producing confidence intervals suffer from a serious deficiency. 
This type of fixed-time statistics is biased against showing the projected peaks of the curves, and thus could obscure the part of the forecast most essential to decision makers.
\begin{figure}
    \centering
    \includegraphics[width=\hsize]{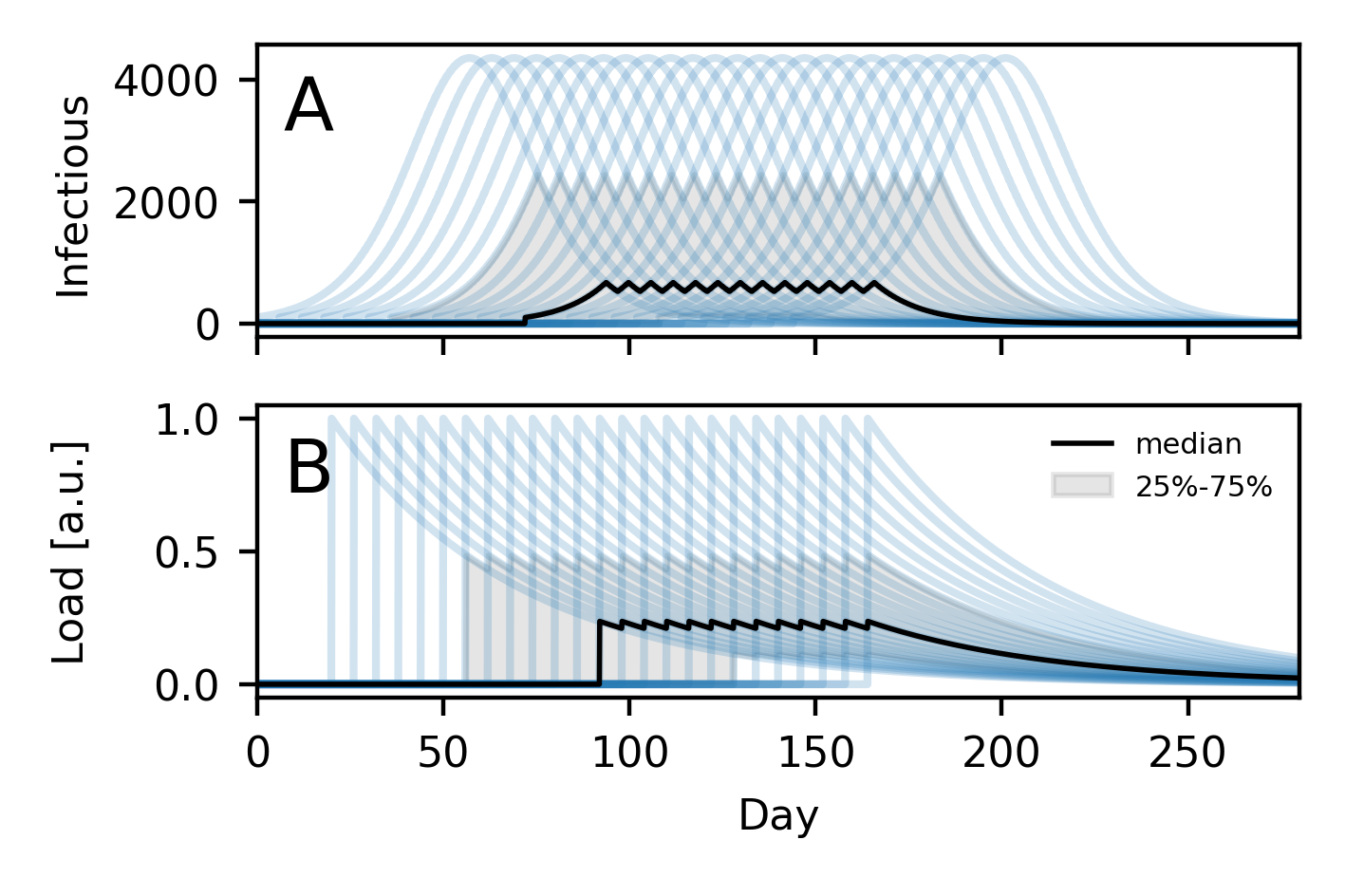}
    \caption{\textbf{Pitfalls in using fixed-time descriptive statistics to summarize ensembles of epidemic curves.} 
    \textbf{A} Simulations of the outbreak on the island Transmithaca (created using a deterministic compartmental model). 
    Blue curves show individual simulations.
    Median and confidence intervals calculated using fixed-time statistics defined in legend.
    Simulations are identical except for the date at which the outbreak starts. 
    The fixed-time descriptive statistics do not capture peak numbers of infections. 
    \textbf{B} Curves of identical shape but with peaks on different days. 
    The fixed-time descriptive statistics only represent the curve ensemble well after day 170 when individual curves take the same shape and do not cross. 
    \label{fig:introducing_problem}
    }
\end{figure}

To illustrate the pitfalls of using fixed-time descriptive statistics to summarize ensembles of simulated epidemic trajectories, let us recount the fictional tale of the inhabitants of \textit{Transmithaca}.

On the island of Transmithaca, one million people lived in complete isolation from the rest of the world. 
A virus had ravaged the outside world, and in the process all viral parameters had become known with perfect precision.
As Transmithaca slowly opened up for outside visitors, the inhabitants knew everything about the virus -- except when it would arrive.
The leaders of Transmithaca asked their epidemiologists to estimate how the disease would impact society.
The epidemiologists simulated a number of scenarios, all with perfect choices of parameters, but different starting dates for the epidemic. 
Their simulations produced an ensemble of epidemic curves and, thinking that the individual simulated epidemic trajectories might clutter the picture, they presented the fixed-time summary statistics shown in grey and black in Fig.~\ref{fig:introducing_problem}A\footnote{Code to reproduce all figures is available at \texttt{www.github.com/jonassjuul/curvestat}. A Python package, \texttt{curvestat}, to produce the curve-based descriptive statistics used in this manuscript can be cloned from {\texttt{ www.github.com/jonassjuul/curvestat}}}. 
Thus, the islanders prepared for an outbreak that might infect between $2\,000$ and $3\,000$ individuals at peak impact.
As we can inspect, however, from the ensemble of time-displaced curves, the actual peak impact in every single case is more than $4\,000$ cases.

The Tragic Tale of Transmithaca is of course a caricature, but it illustrates a fundamental problem in the way ensembles of simulated epidemic trajectories are currently summarized. 
The future course of each curve is decided by the parameters and, crucially, the entire past of the curve.  
These long-term correlations in the shapes of trajectories imply that basing summary statistics on single points in time separately can be misleading.
Because a single curve can enter and leave the marked percentiles, the fixed-time descriptive statistics cannot be expected to capture what a real trajectory might look like. 
In the case of Transmithaca, the summary underestimates the infected-count at peak impact, even though this quantity is identical across all curves. 
This is particularly unfortunate given that -- if you are in charge of pandemic response -- a reliable understanding of peak impact is essential.
Fig.~\ref{fig:introducing_problem}B illustrates that using fixed-time statistics to summarize ensembles can be very inaccurate if curves intersect. The figure also shows that fixed-time statistics are accurate in some special cases, for example when every point on each curve constitutes the same fixed-time percentile for every time step.

\begin{figure}
    \centering
    \includegraphics[]{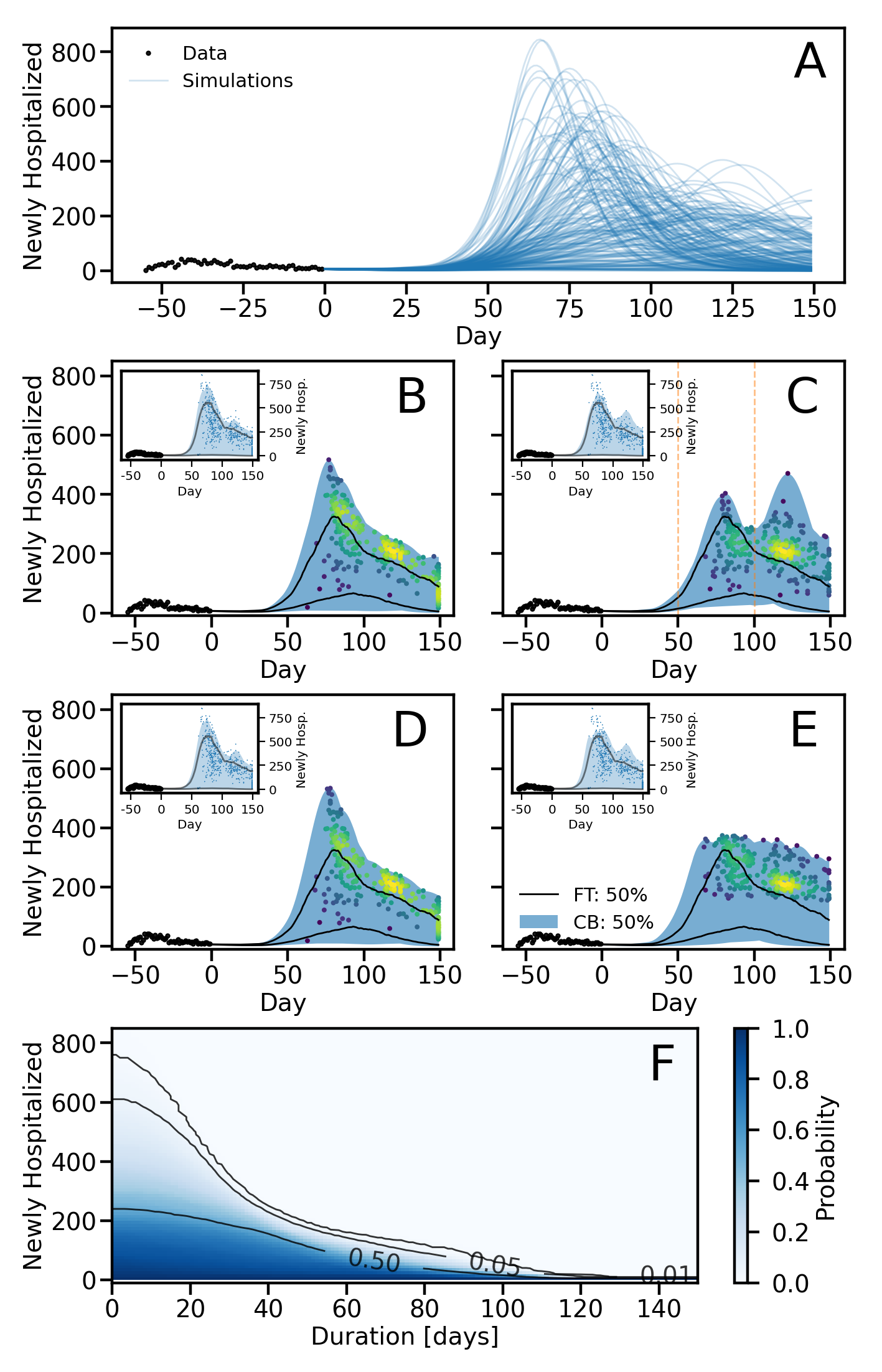} 
    
    \caption{\textbf{Curve-based descriptive statistics to summarize curve ensembles.} 
    \textbf{A} Individual curves from an ensemble of $500$ simulated epidemic trajectories produced as part of the Danish COVID-19 forecasting effort. 
    \textbf{B-E} Curve box plot of the most central $50\%$ (blue) curves plotted alongside the $25$\textsuperscript{th} - $75$\textsuperscript{th} percentiles computed as fixed-time descriptive statistics (black lines). 
    Dots colored using a gaussian kernel density estimator~\cite{scott2015multivariate} shows position and density of peaks of the $50\%$ most central trajectories. Insets show the curve box plot of the most central $90\%$ (light blue) curves plotted alongside the $5$\textsuperscript{th} - $95$\textsuperscript{th} percentiles computed as fixed-time descriptive statistics (grey lines) and a scatter plot showing all trajectory peaks in the ensemble. 
    In \textbf{B-E}, the centrality of curves were ranked in different ways: \textbf{B} All-or-nothing ranking for the full predicted time interval, $N_{\rm curves}=50$ and $N_{\rm samples} = 100$; \textbf{C} All-or-nothing ranking using only the part of curves between day $50$-$100$, $N_{\rm curves}=10$ and $N_{\rm samples} = 100$; \textbf{D} Weighted ranking with a reward $f(t) = e^{-(t-t_0)\ln(2)/7}$ for all curves, $t_0$ being the current date. 
    In this case, the reward for being contained in a sampled envelope gets half as big for every $7$ days.
    \textbf{E} Curves ranked according to their predicted peak number of newly hospitalized patients. 
    The median prediction was deemed most central. 
    \textbf{F} Summarizing the likelihood of certain scenarios as predicted by the curve ensemble. 
    The heatmap color on the point $(x,y)$ indicates the fraction of ensemble curves that at some point predict at least $x$ consecutive days with at least $y$ newly hospitalized patients on each day. 
    \label{fig:Fig2}
    }
\end{figure}

This naturally leads us to the question: Do curves cross in real ensembles of simulated epidemic trajectories? 
We believe that the answer to this question is: Most likely, yes. 
We have been involved in the task of forecasting the epidemic in Denmark, and our sampling of parameters produced an ensemble of trajectories with many such intersections~\cite{ekspertgruppe20maj}.
In Fig.~\ref{fig:Fig2}A we show an ensemble of epidemic trajectories produced as part of the Danish COVID-19 forecasting effort. 
This particular ensemble shows projected daily hospitalizations in the period May 5 - October 1 and was produced under a worst-case assumption that Danes would stop practising social distancing.%
\footnote{Ensembles were produced using a deterministic compartmental model with sampling of parameters based on literature and expert opinions, which is described in detail at \texttt{https://files.ssi.dk/teknisk-gennemgang-af-modellerne-10062020}. The R code used to produce the ensemble is available at \texttt{https://github.com/laecdtu/C19DK}}
For this ensemble, 67\% of curve peaks lie above the 75\textsuperscript{th} fixed-time percentile on the given day. For two other curve ensembles -- which were produced with best-case and moderate assumptions on social distancing, respectively -- 38\% and 54\% of peaks lie above the 75\textsuperscript{th} percentile.
It is clear that percentiles calculated using fixed time statistics (shown using solid black lines) systematically underestimate peak values in these examples. 

As an alternative to using fixed-time statistics in order to recapitulate the desired features of curve ensembles accurately, we propose $2$ alternative summary statistics: 
$1)$ Curve-based descriptive statistics; and $2)$ Summarizing estimated likelihoods of specific scenarios of interest.

\textit{Curve-based descriptive statistics}. 
Whereas fixed-time descriptive statistics separately evaluates the centrality of instantaneous curve values, \textit{curve-based descriptive statistics} rank and visualize centrality of entire curves. 
We suggest using curve box plots to visualize trajectory confidence intervals. 
Curve box plots are sometimes used for summarizing functional ensembles in simulation sciences~\cite{mirzargar2014curve, sun2011functional} and the procedure for constructing a curve box plot is straightforward,
\begin{itemize}
\item[1.] Rank curves from more central to less central,
\item[2.] Plot the envelope containing the most central curves.
\end{itemize}
Here, we define the envelope, $\mathcal{E}(\mathcal{S})$,  of a set of curves, $\mathcal{S}$, as the area spanned by the curves in $\mathcal{S}$. 
More precisely, a point $(t',y)$ is contained by the envelope of $\mathcal{S}$ if there exist curves $c_i(t),c_j(t)\in\mathcal{S}$ such that $c_i(t')\le y \le c_j(t')$.

There are different ways one can rank the centrality of curves, each having its merits. 
With an \emph{all-or-nothing} ranking method all curves start with a centrality score of $0$. $N_{\rm curves}$ curves are drawn uniformly at random from the ensemble and their envelope, $\mathcal{E}_{\rm sample}$, is constructed. 
We then check which ensemble curves are entirely contained by $\mathcal{E}_{\rm sample}$: curve $c_i$ gets $s(c_i)$ added to its centrality score only if all points constituting the curve are contained in $\mathcal{E}_{\rm sample}$. 
The curve-dependent score $s(c_i)$ allows prior information to inform the ranking, e.g.\ a curve’s fit to existing data. 
Uniformly random samples like this are drawn $N_{\rm samples}$ times in total, and each time, centrality scores of all curves are updated. 
In the end, curves with more centrality points are more central in the ensemble. 
In Fig.~\ref{fig:Fig2}B and C we show curve box plots created using all-or-nothing rankings. 
We also show the fixed-time descriptive statistics for the ensemble.

An alternative to the all-or-nothing ranking is what we will call a \textit{weighted} ranking: rewarding curves for each time step they are contained in $\mathcal{E}_{\rm sample}$. 
Again, one draws $N_{\rm curves}$ uniformly randomly from the ensemble. 
Now, however, we add $s(c_i)f(t)$  to curve $c_i$'s centrality score if $c_i$ is contained by $\mathcal{E}_{\rm sample}$ at time $t$. 
The time dependency of the reward can reflect, for example, that some forecasts are expected to be very accurate in the near future but decrease in accuracy with time. Fig.~\ref{fig:Fig2}D shows curve box plots obtained with a time-dependent weighted ranking like this.

In addition to the ranking methods mentioned above, one can rank the curves according to some feature of interest. 
In Fig.~\ref{fig:Fig2}E we show the curve box plots obtained when we rank curves according to their projected maximum values of newly hospitalized cases in a single day; in other words, the median projected peak value received the highest centrality. 

\textit{Likelihoods of specific scenarios of interest}. 
The curve box plots introduced above each visualizes an area that contains a fraction of the ensemble curves. 
Sometimes, however, we may want to go beyond rough estimates of the temporal course of trajectories. 
It might be more interesting to quantify the risk of certain scenarios happening. 
For example, consistent large numbers of hospitalized patients for long periods places a serious burden on healthcare systems~\cite{srinivasan2020rapidly,yang2020clinical}. 
The risk of such scenarios can be explicitly evaluated by counting how often they occur in the ensemble of curves. 
For example, if half of the simulated curves predict that hospitals will get at least $300$ new patients every day for at least $20$ consecutive days, and all curves are considered equally likely, the probability of this scenario is estimated to be 50\%.  
Fig.~\ref{fig:Fig2}F  shows a heatmap communicating this type of risks. 
The color of the point $(x,y)$ in this figure, indicates the fraction of ensemble curves that -- at some point -- have at least $x$ consecutive days with at least $y$ newly hospitalized patients on each day. 
On top of the heatmap, we plot the corresponding contour plot.  
From Fig.~\ref{fig:Fig2}F we directly read off that given this model, the risk of receiving at least $200$ new patients for at least $1$ days in a row is less than $50\%$ and that there is less than $1\%$ risk of receiving at least $400$ new hospitalizations each day for at least $40$ consecutive days. 

If curves are not considered equally probable, but instead each curve, $c_i$, carries a weight, $w(c_i)$, we can generalize the above analysis. In that case, the heatmap value at $(x,y)$ would instead be $\sum_{i\in \mathcal{I}}w(i)/\sum_{j\in\mathcal{S}}w(j)$, where $\mathcal{I}$ is the set of curves that predict at least $x$ consecutive days with at least $y$ new patients hospitalized on each day.

In summary, when making forecasts of epidemic trajectories, it is important to represent the resulting curve ensembles in a way that captures the quantities of interest in an intuitive way. 
Here we have argued that computing confidence intervals using fixed-time descriptive systematically suppresses trajectory extremes. 
This is natural. 
Fixed-time descriptive statistics are designed to show the least extreme predictions on a given date, not to take entire curves into account. 
In a situation where the projected peak numbers of hospitalized patients are of the utmost importance to decision makers and the public, however, this is unfortunate. 
We hope that this paper raises awareness of this pitfall of fixed-time descriptive statistics for summarizing ensembles of epidemiological trajectories. 
In addition to identifying this shortcoming, we have suggested how curve ensemble extremes can be summarized and visualized instead. 

The methods we have suggested here -- plotting curve box plots and estimating likelihoods of specific scenarios of interest -- focus on ensemble trajectories instead of single time steps. 
This approach provides forecasters with statistical tools that do not filter out curve extremes. 
At the same time, however, it is clear that there is still important research to be done. 
We have suggested different ways of ranking centrality of ensemble curves. 
Each ranking has its merits: For example, the all-or-nothing ranking penalizes curves which are shaped differently than other ensemble curves and a weighted ranking can reward curves that are central in the very near future. 
A thorough investigation of these merits and trade-offs is needed. 
Until this has been investigated, we encourage researchers to be creative and mindful about the problems that each statistical method could have. 
And to communicate this uncertainty openly to decision makers.

\section*{Acknowledgments}
The authors are thankful to the members of the SSI COVID-19 modeling group for an excellent collaboration and to Carl T. Bergstrom for comments on an early version of the manuscript. J.L.J and S.L. received additional funding through the HOPE project (Carlsberg Foundation).

\section*{Author contributions}
J.L.J. and S.L. conceived the idea. J.L.J. performed simulations, analysis and calculations. K.G. and L.E.C. devised and performed epidemiological simulations. All authors contributed to discussions and wrote the manuscript.

\bibliography{visualizing_ensembles_of_curves.bib}

\end{document}